\documentclass[showpacs,10pt,twocolumn,prb]{revtex4-1}

\usepackage{amsmath}
\usepackage{amssymb}
\usepackage{graphicx}
\usepackage{amssymb}
\usepackage{graphics}
\usepackage{epsfig}
\usepackage{CJK}
\usepackage{color}

\begin{document}

\begin{CJK*}{GBK}{Song}
\title{Critical behavior of quasi-two-dimensional weak itinerant ferromagnet trigonal chromium telluride Cr$_{0.62}$Te}
\author{Yu Liu and C. Petrovic}
\affiliation{Condensed Matter Physics and Materials Science Department, Brookhaven National Laboratory, Upton, New York 11973, USA}
\date{\today}

\begin{abstract}
The critical properties of flux-grown single-crystalline quasi-two-dimensional weak itinerant ferromagnet Cr$_{0.62}$Te were investigated by bulk dc magnetization around the paramagnetic (PM) to ferromagnetic (FM) phase transition. Critical exponents $\beta = 0.315(7)$ with a critical temperature $T_c = 230.6(3)$ K and $\gamma = 1.81(2)$ with $T_c = 229.1(1)$ K are obtained by the Kouvel-Fisher method whereas $\delta = 6.35(4)$ is obtained by a critical isotherm analysis at $T_c = 230$ K. With these obtained exponents, the magnetization-field-temperature curves collapse into two independent curves following a single scaling equation $M|\frac{T-T_c}{T_c}|^{-\beta} = f_\pm(H|\frac{T-T_c}{T_c}|^{-\beta\delta})$ around $T_c$, suggesting the reliability of the obtained exponents. Additionally, the determined exponents of Cr$_{0.62}$Te exhibit an Ising-like behavior with a change from short-range order to long-range order in the nature of magnetic interaction and with an extension from 2D to 3D on cooling through $T_c$.
\end{abstract}

\maketitle
\end{CJK*}

\section{INTRODUCTION}

Two-dimensional (2D) materials such as graphene and ultrathin transition-metal dichalcogenides have been attracting significant interest due to their highly tunable physical properties and immense potential in scalable device applications.\cite{Geim, Chhowalla, Hu, Bhimanapati} However, in contrast to mechanical and optoelectronic properties, the investigation of magnetism in 2D materials has received little attention. Recently, the Cr-chalcogenides are of great interest for possible application in spintronic technologies. The CrXTe$_3$ (X = Si, Ge, Sn) have been identified as promising candidate for long-range magnetism in monolayer.\cite{Sivadas, Zhuang1, Lin1} CrSiTe$_3$ exhibits ferromagnetic (FM) ordering at $\sim$ 32 K in the bulk,\cite{Casto} and it can be enhanced to $\sim$ 80 K in monolayer and few-layer samples.\cite{Lin2} Bulk CrGeTe$_3$ are ferromagnetic at $\sim$ 61 K, which is still somewhat low for spintronic applications.\cite{Zhang2}

In the binary Cr-based chalcogenides, tellurides CrTe, Cr$_3$Te$_4$, Cr$_2$Te$_3$, Cr$_5$Te$_8$ are ferromagnetic with metallic conductivity,\cite{Herbert, Street, Hamasaki, Akram, Lukoschus, Huang1, Huang2} while selenides CrSe, Cr$_3$Se$_4$, Cr$_2$Se$_3$, Cr$_5$Se$_8$ and sulfides CrS, Cr$_5$S$_6$, Cr$_3$S$_4$, Cr$_2$S$_3$, Cr$_5$S$_8$ are predominantly antiferromagnetic (AFM) showing either metallic or semiconducting behavior.\cite{Chevreton, Adachi, Li, Yuzuri, Vaqueiro} Among these compounds, Cr$_{1-x}$Te system shows FM with $T_c$ of 170 $\sim$ 360 K.\cite{Herbert} Cr$_{1-x}$Te with $x < 0.1$ crystallize in the hexagonal NiAs structure, while Cr$_3$Te$_4$ ($x = 0.25$) and Cr$_2$Te$_3$ ($x = 0.33$) form monoclinic and trigonal crystal structures where Cr vacancies occupy in every second metal layer. According to neutron-diffraction studies, the saturation magnetization in this system is small and it could be partly explained if spin canting is taken into consideration for $x =$ 0.125, 0.167, and 0.25.\cite{Andresen} The magnetic moment induced on the Cr ion for $x = 0.25$ is close to an integral number of Bohr magnetons, suggesting the existence of mixed valence Cr.\cite{Andresen} However, for Cr$_2$Te$_3$ ($x = 0.33$), the ordered magnetic moment of 2.65 $\sim$ 2.70 $\mu_B$ deduced from the neutron diffraction is much smaller than that calculated using the ionic model 3 $\mu_B$, suggesting the itinerant nature of the $d$ electrons.\cite{Hamasaki, Andresen} The electron correlation effects in itinerant ferromagnets has also been discussed in the photoemission spectra.\cite{Shimada} Until now, only a few studies were performed on Cr$_{1-x}$Te with $x = 0.375$, of which the trigonal phase exhibits a higher $T_c$ than that of the monoclinic phase.\cite{Lukoschus}

Trigonal Cr$_{0.62}$Te exhibits weak itinerant ferromagnetic character with $T_c \approx 237$ K. In order to understand the nature of the FM transition, we investigated its critical behavior by modified Arrott plot, Kouvel-Fisher plot, and critical isotherm analysis. The determined exponents $\beta$ = 0.315(7) with $T_c$ = 230.65(26) K, $\gamma$ = 1.81(2) with $T_c$ = 229.11(5) K, and $\delta$ = 6.35(4) at $T_c$ = 230 K indicate a change in the nature of magnetic interaction passing from short-range order to long-range order with an extension from 2D to 3D on cooling through $T_c$.

\section{EXPERIMENTAL DETAILS}

The trigonal Cr$_{0.62}$Te single crystals were grown by the self-flux technique starting from an intimate mixture of pure elements Cr (99.99 $\%$, Alfa Aesar) powder and Te (99.9999 $\%$, Alfa Aesar) pieces with a molar ratio of 0.06 : 0.94. Starting materials were sealed in an evacuated quartz tube, which was heated to 900 $^\circ$C over 20 h, held at 900 $^\circ$C for 3 h, and then slowly cooled to 500 $^\circ$C at a rate of 1 $^\circ$C/h. X-ray diffraction (XRD) data were taken with Cu K$_{\alpha}$ ($\lambda=0.15418$ nm) radiation of Rigaku Miniflex powder diffractometer. The element analysis was performed using an energy-dispersive x-ray spectroscopy (EDX) in a JEOL LSM-6500 scanning electron microscope, confirming a near-stoichiometric Cr$_{0.62}$Te single crystal. The magnetization was measured in a Quantum Design Magnetic Property Measurement System (MPMS-XL5). The isothermal $M(H)$ curves are measured in $\Delta T$ = 1 K intervals. The applied magnetic field ($H_a$) has been corrected for the internal field as $H = H_a - NM$, where $M$ is the measured magnetization and $N$ is the demagnetization factor. The corrected $H$ was used for the analysis of critical behavior.

\section{RESULTS AND DISCUSSIONS}

\begin{figure}
\centerline{\includegraphics[scale=1.0]{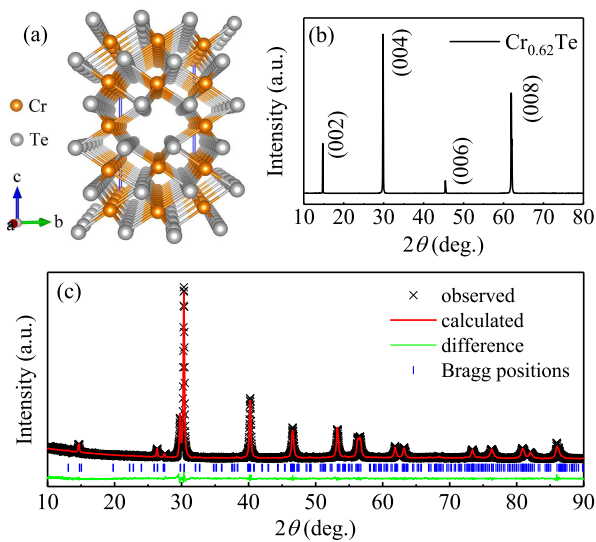}}
\caption{(Color online). (a) Crystal structure of Cr$_{0.62}$Te. (b) Single-crystal x-ray diffraction (XRD) and (c) powder XRD pattern of Cr$_{0.62}$Te. The vertical tick marks represent Bragg reflections of the $P\bar{3}m1$ space group.}
\label{XRD}
\end{figure}

\begin{figure}
\centerline{\includegraphics[scale=0.95]{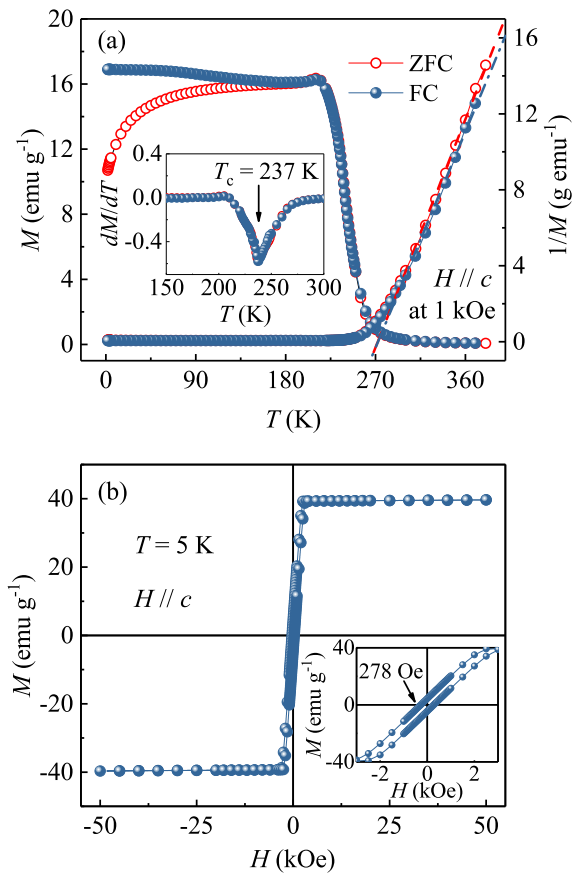}}
\caption{(Color online). (a) Temperature dependence of magnetization for Cr$_{0.62}$Te measured in the external magnetic field $H$ = 1 kOe applied along the $c$ axis with zero-field-cooling (ZFC) and field-cooling (FC) modes. The dashed lines are fits by the modified Curie-Weiss law $\chi = \frac{C}{T-\theta}+\chi_0$, where $\chi_0$ is the temperature-independent susceptibility, $C$ is the Curie-Weiss constant, and $\theta$ is the Weiss temperature. Inset: the derivative magnetization $dM/dT$ vs $T$. (b) Field dependence of magnetization for Cr$_{0.62}$Te measured at $T$ = 5 K. Inset: the magnification of the low field region.}
\label{MTH}
\end{figure}

\begin{figure}
\centerline{\includegraphics[scale=0.9]{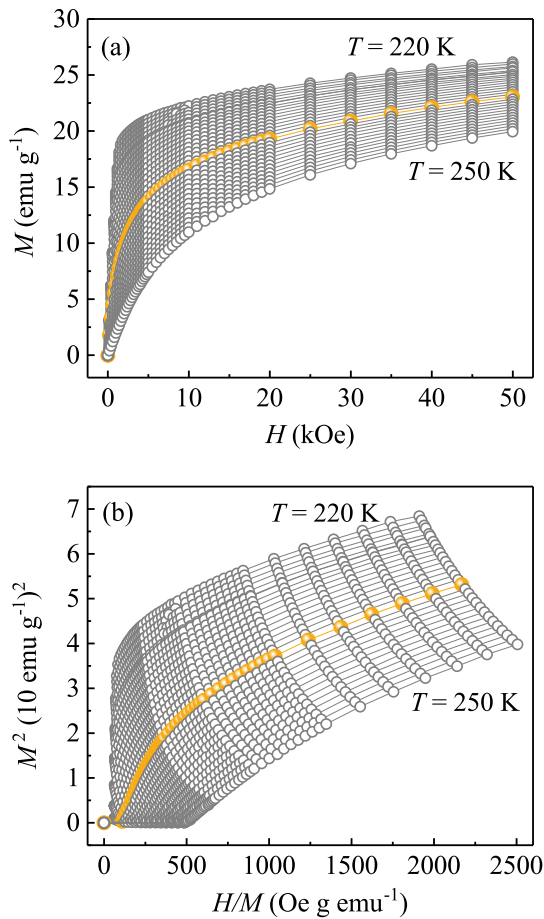}}
\caption{(Color online). (a) Typical initial isothermal magnetization curves measured along the $c$ axis around $T_c$ = 237 K (in an orange symbol and line) for Cr$_{0.62}$Te. (b) Arrott plots of $M^2$ vs $H/M$ around $T_c$ for Cr$_{0.62}$Te.}
\label{Arrot}
\end{figure}

The Cr-Te phase diagram was summarized by Herbert \emph{et al.}.\cite{Herbert} The multiple Cr$_{1-x}$Te phases with structures intermediate between the NiAs and the CdI$_2$ types depend on the different arrangements of metal atom vacancies. So far, the trigonal Cr$_{0.62}$Te in which the Cr atoms are located on four crystallographically different sites leading to the formation of a five-layer superstructure of the CdI$_2$ type [Fig. 1(a)] has been rarely studied.\cite{Bensch} In the CdI$_2$-type structure the second basal planes for Cr atoms of NiAs-type structure are partially occupied by Cr vacancies and the Cr atoms are surrounded by the octahedral environment of Te anions.\cite{Lukoschus} Figure 1(b) presents the single crystal x-ray diffraction (XRD) pattern of Cr$_{0.62}$Te. Only $(00l)$ peaks are detected, indicating the crystal surface is normal to the $c$ axis with the plate-shaped surface parallel to the $ab$-plane. As shown in Fig. 1(c), the powder XRD pattern are well fitted with the $P\bar{3}m1$ space group. The determined lattice parameters $a = 0.7792(2)$ nm and $c = 1.1980(2)$ nm are very close to the values in previous report.\cite{Lukoschus}

Figure 2(a) shows the temperature dependence of magnetization $M(T)$ measured in $H$ = 1 kOe applied parallel to the $c$ axis, in which a clear paramagnetic (PM) to ferromagnetic (FM) transition is observed. As shown in the inset in Fig. 2(a), the critical temperature $T_c \approx 237$ K is roughly determined from the minimum of the derivative $dM/dT$ curve, which is in good agreement with the value reported previously.\cite{Lukoschus} The zero-field-cooling (ZFC) and field-cooling (FC) curves show significant splitting at low temperatures, indicating a strong magnetocrystalline anisotropy or spin canting. The $1/M$ vs $T$ is also plotted in Fig. 2(a). A linear fit in the high temperature range of 290 $\sim$ 350 K yields the Weiss temperature $\theta = 272(1)$ K, indicating predominant FM exchange interaction. The deduced effective moment $\mu_{\textrm{eff}}$ = 3.93(3) $\mu_B/$Cr for the ZFC curve is close to $\mu_{\textrm{eff}}$ = 4.04(5) $\mu_B/$Cr for the FC data, which is in agreement with the theoretical value expected for Cr$^{3+}$ of 3.87 $\mu_B$. Figure 2(b) displays the isothermal magnetization measured at $T$ = 5 K. The saturation field is $H_s \approx 3 $ kOe and the saturation moment at $T$ = 5 K is $M_s \approx$ 1.814(1) $\mu_B/$Cr. The inset in Fig. 2(b) shows the $M(H)$ in the low field region and little hysteresis with the coercive force $H_c = 278$ Oe. All these results are in good agreement with the previous report.\cite{Lukoschus} Then we calculated the Rhodes-Wohlfarth ratio (RWR) for Cr$_{0.62}$Te, which is defined as $P_c/P_s$ with $P_c$ obtained from the effective moment $P_c(P_c+2) = P_{eff}^2$ and $P_s$ is the saturation moment obtained in the ordered state.\cite{Wohlfarth, Moriya} RWR is 1 for localized systems and is larger in an itinerant system. Here we obtain RWR = 1.69 for ZFC and RWR = 1.74 for FC, indicating weak itinerant character of FM in Cr$_{0.62}$Te.

\begin{figure*}
\centerline{\includegraphics[scale=0.9]{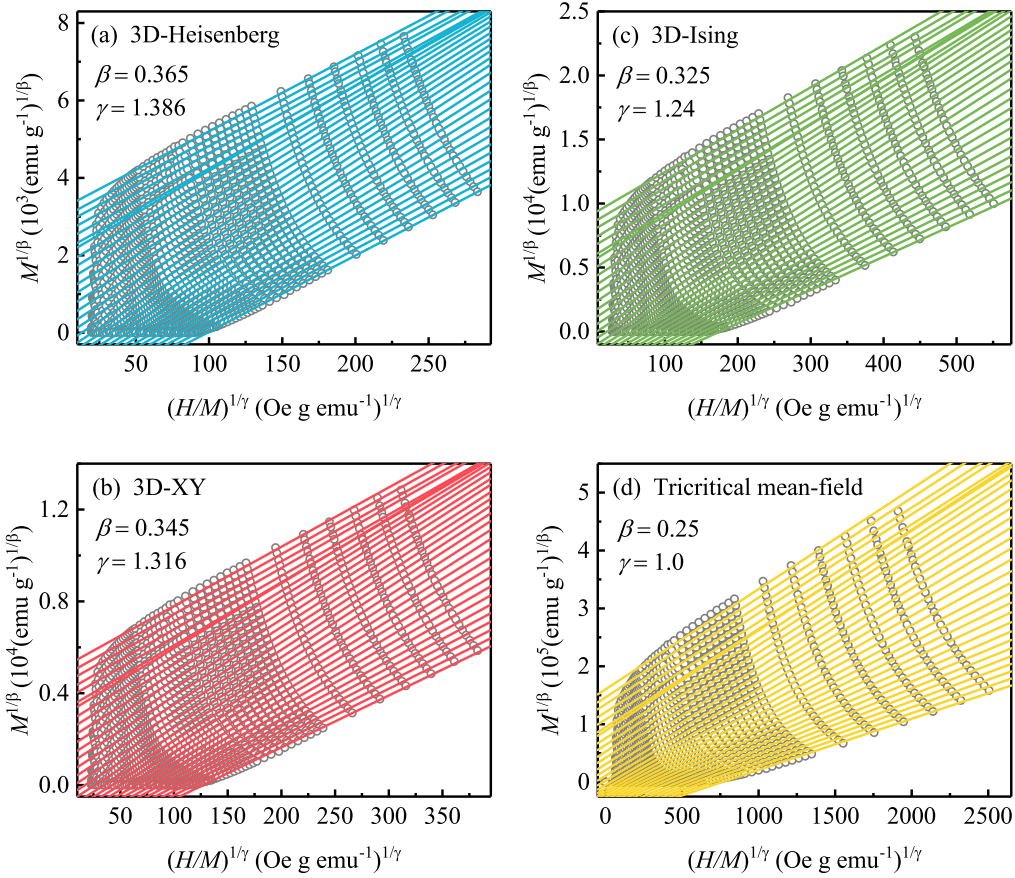}}
\caption{(Color online). The isotherms of $M^{1/\beta}$ vs $(H/M)^{1/\gamma}$ with parameters of (a) 3D Heisenberg model, (b) 3D XY model, (c) 3D Ising model, and (d) tricritical mean-field model. The straight lines are the linear fit of isotherms at different temperatures.}
\label{3D}
\end{figure*}

In order to understand the nature of the FM transition in Cr$_{0.62}$Te, one approach is to study in detail the critical exponents associated with the transition. Isothermal magnetization $M(H)$ around $T_c$ was measured from $T$ = 220 K to $T$ = 250 K at intervals of 1 K to investigate the critical behavior, as shown in Fig. 3(a). Figure 3(b) presents the Arrott plot of $M^2$ vs $H/M$. Generally, $M^2$ vs $H/M$ should be a series of parallel straight lines in the high field range in the Arrott plot.\cite{Arrott1} The intercept of the $M^2$ as a function of $H/M$ on the $H/M$ axis is negative above $T_c$ while it is positive below $T_c$. The line of $M^2$ vs $H/M$ at $T_c$ should pass through the origin. According to the criterion proposed by Banerjee,\cite{Banerjee} the order of the magnetic transition can be determined from the slope of the straight line: the positive slope corresponds to the second order transition while the negative to the first order one. Apparently, the positive slope of the $M^2$ vs $H/M$ implies that the PM-FM transition in Cr$_{0.62}$Te is a second order one, as shown in Fig. 3(b). However, all the curves in this plot are nonlinear and show downward curvature even in the high field region, which indicates that the long-range Landau mean-field theory with $\beta$ = 0.5 and $\gamma$ = 1.0 is not satisfied for Cr$_{0.62}$Te according to Arrot-Noaks equation of state $(H/M)^{1/\gamma} = a\varepsilon+bM^{1/\beta}$, where $\varepsilon = (T-T_c)/T_c$ is the reduced temperature, $a$ and $b$ are constants.\cite{Arrott2} Hence, a modified Arrott plot should be used to obtain the critical parameters.

It is well known that for a second-order transition, its critical behavior can be characterized in detail by a series of interrelated critical exponents.\cite{Stanley} In the vicinity of a second-order phase transition, the divergence of correlation length $\xi = \xi_0 |(T-T_c)/T_c|^{-\nu}$ leads to universal scaling laws for the spontaneous magnetization $M_s$ and the inverse initial magnetic susceptibility $\chi_0^{-1}$. The mathematical definitions of the exponents from magnetization can be described as:
\begin{equation}
M_s (T) = M_0(-\varepsilon)^\beta, \varepsilon < 0, T < T_c
\end{equation}
\begin{equation}
\chi_0^{-1} (T) = (h_0/m_0)\varepsilon^\gamma, \varepsilon > 0, T > T_c
\end{equation}
\begin{equation}
M = DH^{1/\delta}, \varepsilon = 0, T = T_c
\end{equation}
where $\varepsilon = (T-T_c)/T_c$ is the reduced temperature, and $M_0$, $h_0/m_0$ and $D$ are the critical amplitudes.\cite{Stanley, Fisher} Parameters $\beta$ (associated with $M_s$), $\gamma$ (associated with $\chi_0$), and $\delta$ (associated with $T_c$) are critical exponents. 3D Heisenberg model ($\beta = 0.365, \gamma = 1.386$), 3D XY model ($\beta = 0.345, \gamma = 1.316$), 3D Ising model ($\beta = 0.325, \gamma = 1.24$), and tricritical mean-field model ($\beta = 0.25, \gamma = 1.0$) are used to construct a modified Arrott plots,\cite{Zhang3} as given in Fig. 4. All these models yield quasi-straight lines in high field region, however, the lines in Fig. 4(d) are not parallel to each other, indicating that the tricritical mean-field model is not satisfied. In order to distinguish which model is the best, normalized slopes ($NS$), which is defined as $NS = S(T)/S(T_c)$ (where $S(T)$ is the slope of $M^{1/\beta}$ vs $(H/M)^{1/\gamma}$), are plotted in Fig. 5. In an ideal model, all values of $NS$ should be equal to 1.0 because the modified Arrott plot should consist of a series of parallel straight lines. For Cr$_{0.62}$Te, the $NS$ of 3D Ising model is close to $NS = 1$ mostly below $T_c$, while that of 3D Heisenberg model is the best above $T_c$, indicating that the critical behavior of Cr$_{0.62}$Te may not belong to a single universality class.

\begin{figure}
\centerline{\includegraphics[scale=0.94]{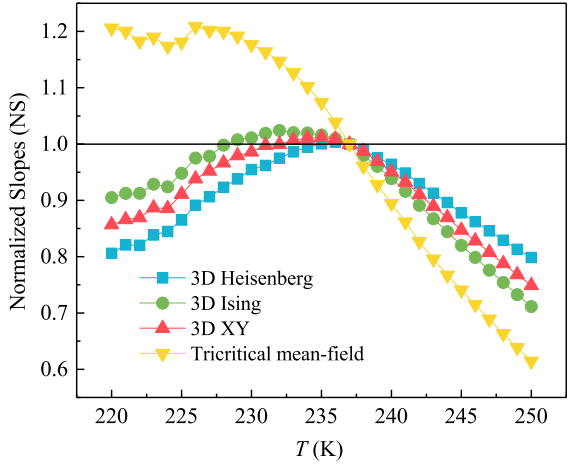}}
\caption{(Color online). Temperature dependence of the normalized slopes $NS = S(T)/S(T_c)$.}
\label{NS}
\end{figure}

\begin{figure}
\centerline{\includegraphics[scale=1]{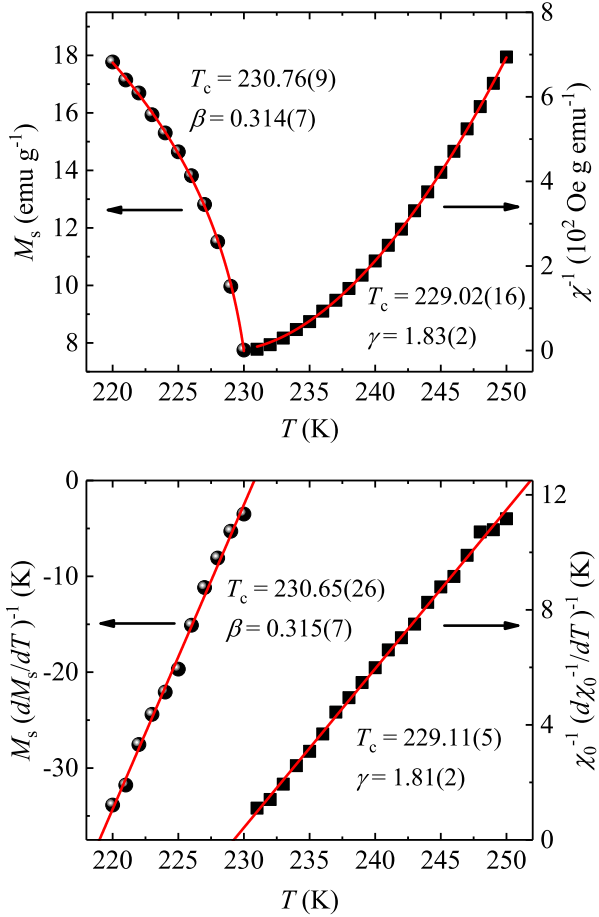}}
\caption{(Color online). (a) Temperature dependence of the spontaneous magnetization $M_s$ (left) and the inverse initial susceptibility $\chi_0^{-1}$ (right) with solid fitting curves for Cr$_{0.62}$Te. (b) Kouvel-Fisher plots of $M_s(dM_s/dT)^{-1}$ (left) and $\chi_0^{-1}(d\chi_0^{-1}/dT)^{-1}$ (right) with solid fitting curves for Cr$_{0.62}$Te.}
\label{KF}
\end{figure}

In order to obtain the precise critical exponents $\beta$ and $\gamma$, a rigorous iterative method has been used.\cite{Pramanik} The linear extrapolation from the high field region to the intercepts with the axis $M^{1/\beta}$ and $(H/M)^{1/\gamma}$ yields reliable values of $M_s(T)$ and $\chi_0^{-1}(T)$. A set of $\beta$ and $\gamma$ can be obtained by fitting data following the Eqs (1) and (2), which is used to reconstruct a new modified Arrott plot. Then, a new set of $\beta$ and $\gamma$ can be obtained. This procedure was repeated until the values of $\beta$ and $\gamma$ do not change. By this method, the obtained critical exponents do not depend on the initial parameters, which confirms these critical exponents are reliable and intrinsic. Figure 6(a) presents the final $M_s(T)$ and $\chi_0^{-1}(T)$ with the solid fitting curves. The critical exponents $\beta = 0.314(7)$ with $T_c = 230.76(9)$ K and $\gamma = 1.83(2)$ with $T_c = 229.0(2)$ K are obtained. More accurately, the critical exponents can be determined according to the Kouvel-Fisher (KF) method:\cite{Kouvel}
\begin{equation}
\frac{M_s(T)}{dM_s(T)/dT} = \frac{T-T_c}{\beta}
\end{equation}
\begin{equation}
\frac{\chi_0^{-1}(T)}{d\chi_0^{-1}(T)/dT} = \frac{T-T_c}{\gamma}
\end{equation}
According to the KF method, $M_s(T)/[dM_s(T)/dT]$ and $\chi_0^{-1}(T)/[d\chi_0^{-1}(T)/dT]$ are linear functions of temperature with slopes of $1/\beta$ and $1/\gamma$, respectively. As shown in Fig. 6(b), the linear fits give $\beta = 0.315(7)$ with $T_c = 230.6(3)$ K and $\gamma = 1.81(2)$ with $T_c = 229.1(1)$ K, respectively, which are consistent with those generated by the modified Arrott plot.

\begin{figure}
\centerline{\includegraphics[scale=0.895]{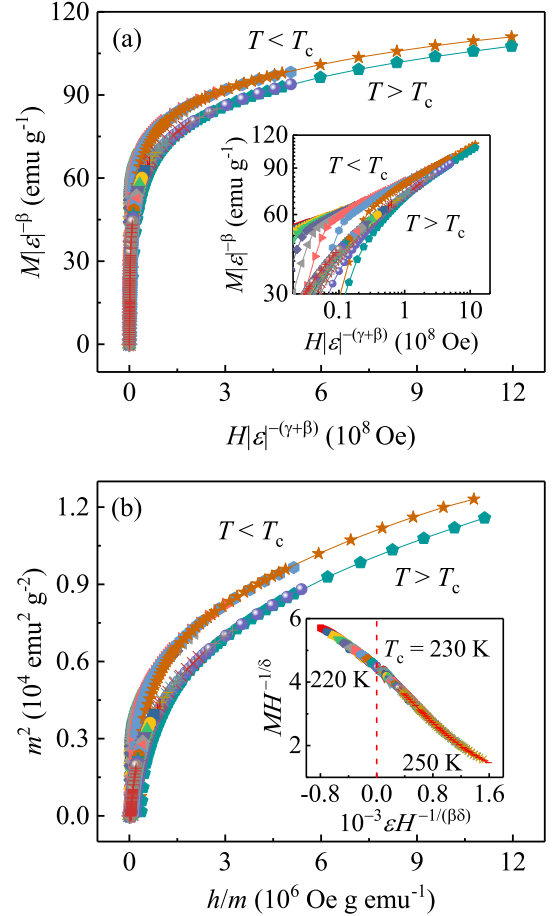}}
\caption{(Color online). (a) Scaling plots of renormalized magnetization $m$ versus renormalized field $h$ below and above $T_c$ for Cr$_{0.62}$Te. Inset: the same plots in log-log scale. (b) the renormalized magnetization and field replotted in the form of $m^2$ versus $h/m$ for Cr$_{0.62}$Te. Inset: the rescaling of the $M(H)$ curves by $MH^{-1/\delta}$ vs $\varepsilon H^{-1/(\beta\delta)}$.}
\label{magnetism}
\end{figure}

As confirmation, the critical exponents can be tested according to the prediction of the scaling hypothesis. In the critical asymptotic region, the magnetic equation can be written as:\cite{Stanley}
\begin{equation}
M(H,\varepsilon) = \varepsilon^\beta f_\pm(H/\varepsilon^{\beta+\gamma})
\end{equation}
where $f_+$ for $T>T_c$ and $f_-$ for $T<T_c$, respectively, are the regular functions. In terms of renormalized magnetization $m\equiv\varepsilon^{-\beta}M(H,\varepsilon)$ and renormalized field $h\equiv\varepsilon^{-(\beta+\gamma)}H$, the Eq.(6) can be written as:
\begin{equation}
m = f_\pm(h)
\end{equation}
it implies that for true scaling relations and right choice of $\beta$, $\gamma$, and $\delta$ values, scaled $m$ and $h$ will fall on two universal curves: one above $T_c$ and another below $T_c$. This is an important criterion for the critical regime. Following Eq. (7), scaled $m$ vs scaled $h$ has been plotted in Fig. 7(a), with the logarithmic scale in the inset of Fig. 7(a). It is rather significant that all the data collapse into two separate branches: one below $T_c$ and another above $T_c$. The reliability of the exponents and $T_c$ has been further ensured with more rigorous method by plotting $m^2$ vs $h/m$, as shown in Fig. 7(b), where all data also fall on two independent branches. This clearly indicates that the interactions get properly renormalized in critical regime following scaling equation of state. In addition, the scaling equation of state takes another form:
\begin{equation}
\frac{H}{M^\delta} = k(\frac{\varepsilon}{H^{1/\beta}})
\end{equation}
where $k(x)$ is the scaling function. Based on Eq. (8), all experimental curves will collapse into a single curve. The inset in Fig. 7(b) shows the $MH^{-1/\delta}$ vs $\varepsilon H^{-1/(\beta\delta)}$ for Cr$_{0.62}$Te, where the experimental data collapse into a single curve, and $T_c$ locates at the zero point of the horizontal axis. The well-rescaled curves further confirm the reliability of the obtained critical exponents.

\begin{figure}
\centerline{\includegraphics[scale=0.94]{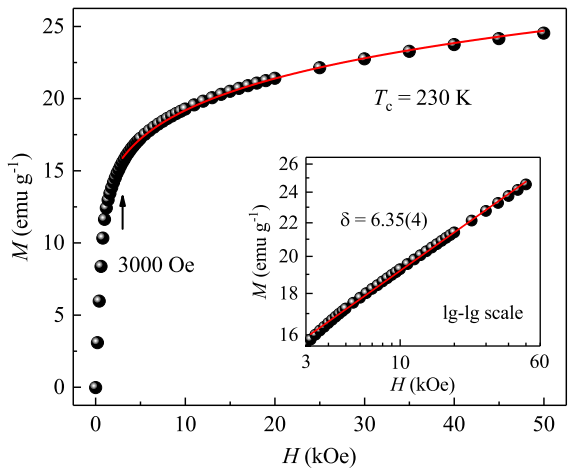}}
\caption{(Color online). Isotherm $M$ vs $H$ plot collected at $T_c$ = 230 K for Cr$_{0.62}$Te. Inset: the same plot in log-log scale with a solid fitting curve.}
\label{MH}
\end{figure}

The critical exponent $\delta$ can be determined by the critical isotherm analysis from $M(H)$ at $T_c$ following Eq. (3). It is determined $T_c = 230$ K from the obtained critical exponents. Thus, the isothermal magnetization $M(H)$ at $T_c$ = 230 K is shown in Fig. 8, where the inset plots the logarithmic scale. The log($M$)-log($H$) relation yields straight line at higher field range ($H > 3$ kOe) with the slope $1/\delta$, where $\delta$ = 6.35(4) is obtained. According to the statistical theory, these three critical exponents should agree with the Widom scaling relation:\cite{Kadanoff}
\begin{equation}
\delta = 1+\frac{\gamma}{\beta}
\end{equation}
Using the $\beta$ and $\gamma$ values determined from Modified Arrott plot and Kouvel-Fisher plot, we obtain $\delta$ = 6.83(7) and $\delta$ = 6.75(6), respectively, which slightly deviates from that obtained from critical isotherm analysis $\delta$ = 6.35(4). The slight deviation from the Widom scaling law indicates that the critical behavior of Cr$_{0.62}$Te does not belong to a single university class as a result of possible complex competition of several magnetic interactions.

\begin{table*}
\caption{\label{tab1}Comparison of critical exponents of Cr$_{0.62}$Te with different theoretical models.}
\begin{ruledtabular}
\begin{tabular}{lllllll}
  Composition & Reference & Technique & $T_c$ & $\beta$ & $\gamma$ & $\delta$ \\
  \hline
  Cr$_{0.62}$Te & This work & Modified Arrott plot & 230.76(9) & 0.314(7) & 1.83(2) & 6.83(7) \\
  & This work & Kouvel-Fisher plot & 230.65(26) & 0.315(7) & 1.81(2) & 6.75(6) \\
  & This work &Critical isotherm  &   &   &   & 6.35(4) \\
  \{d:n\} = \{2:1\} &  &  & &  &  & \\
  2D short-range ($J(r)\sim e^{-r/b}$) & 37 & Theory & & 0.125 & \textbf{1.75} & 15.0 \\
  \{d:n\} = \{2:1\} &  &  & &  &  & \\
  2D long-range ($J(r)\sim r^{-(d+\sigma)}$) & 37 & Theory & & \textbf{0.298} & 1.393 & 5.67 \\
  \{d:n\} = \{3:1\} &  &  & &  &  & \\
  3D Ising model & 38 & Theory & & \textbf{0.325} & 1.24 & 4.82 \\
  \{d:n\} = \{3:2\} &  &  & &  &  & \\
  3D XY model & 38 & Theory & & 0.345 & 1.316 & 4.81 \\
  \{d:n\} = \{3:3\} &  &  & &  &  & \\
  3D Heisenberg model & 38 & Theory & & 0.365 & 1.386 & 4.8 \\
  Mean-field model & 38 & Theory & & 0.5 & 1.0 & 3.0\\
  Tricritical mean field & 39 & Theory & & 0.25 & 1.0 & 5.0
\end{tabular}
\end{ruledtabular}
\end{table*}

The obtained critical exponents of Cr$_{0.62}$Te, as well as those of different theoretical models,\cite{Fisher1972, Kaul, Huang3} are listed in Table I for comparison. As we can see, the critical exponents of Cr$_{0.62}$Te do not belong to any single universality class. For a homogeneous magnet, the universality class of the magnetic phase transition depends on the exchange distance $J(r)$. A renormalization group theory analysis suggests that the long-range exchange interaction decays as $J(r) \sim r^{-(d+\sigma)}$, where $d$ is the spatial dimension and $\sigma$ is a positive constant, and the short-range exchange interaction decays as $J(r) \sim e^{-r/b}$, where $b$ is the spatial scaling factor.\cite{Fisher, Fisher1972, LeGuillou} The long or short range of spin interaction depends on the value of $\sigma$, which is determined by:\cite{Fisher, Fischer, Pramanik}
\begin{multline}
\gamma = 1+\frac{4}{d}(\frac{n+2}{n+8})\Delta\sigma+\frac{8(n+2)(n-4)}{d^2(n+8)^2}\\\times[1+\frac{2G(\frac{d}{2})(7n+20)}{(n-4)(n+8)}]\Delta\sigma^2
\end{multline}
where $\Delta\sigma = (\sigma-\frac{d}{2})$ and $G(\frac{d}{2})=3-\frac{1}{4}(\frac{d}{2})^2$, $n$ is the spin dimensionality. The short-range Heisenberg model is valid when $\sigma > 2$, while the long-range mean-field model is satisfied when $\sigma < 3/2$. In present case, based on the experimental $\gamma = 1.81(2)$, it can be obtained that $\sigma = 1.626$ following Eq. (10). It can be seen that $\sigma$ lies between the long-range and short-range interaction ($3/2 < \sigma < 2$). In addition, one can see that $\gamma = 1.81(2)$ is close to the short-range interaction with $\{d:n\} = \{2:1\}$ ($\beta = 0.125$, $\gamma = 1.75$, and $\delta = 15.0$), however, $\beta = 0.315(7)$ approaches to the long-range interaction with $\{d:n\} = \{2:1\}$ ($\beta = 0.298$, $\gamma = 1.392$, and $\delta = 5.67$) and/or 3D Ising model with $\{d:n\} = \{3:1\}$ ($\beta = 0.325$, $\gamma = 1.24$, and $\delta = 4.82$) (as listed in Table I). The previous photoemission spectrum of Cr$_{0.62}$Te also confirmed the existence of short-range FM interaction above $T_c$.\cite{Shimada} Here the extension from $d = 2$ to $d = 3$ passing through $T_c$ indicates that Cr$_{0.62}$Te features quasi-two-dimensional character of chemical bonds and non-negligible exchange interaction along the c axis through the Cr deficient layers; the $n = 1$ generally implies uniaxial or Ising-like magnetic interaction. In addition, it is found that the experimental critical exponent of $\gamma$ is slightly larger than the theoretical value of 2D short-range interaction, and the value of $\beta$ lies between the theoretical values of 2D long-range interaction and 3D Ising model, further indicating non-negligible interlayer coupling and strong electron-electron correlation in this material.\cite{Lukoschus, Shimada} The significant hybridization between Cr $3d$ and Te $5p$ bands and strong electron-correlation effects were also previously observed by photoemission spectroscopy,\cite{Shimada} confirming the weak itinerant ferromagnetic character of Cr$_{0.62}$Te.

\section{CONCLUSIONS}

In summary, we have made a comprehensive study of the critical region at the PM-FM phase transition in the weak itinerant ferromagnet Cr$_{0.62}$Te. This transition is identified to be second order in nature. The critical exponents $\beta$, $\gamma$, and $\delta$ estimated from various techniques match reasonably well and follow the scaling equation, confirming that the obtained exponents are unambiguous and intrinsic to the material. Above $T_c$ the critical exponent $\gamma = 1.81(2)$ is close to the short-range interaction with $\{d:n\} = \{2:1\}$ ($\beta = 0.125$, $\gamma = 1.75$, and $\delta = 15.0$) whereas below $T_c$ the critical exponent $\beta = 0.315(7)$ lies between the long-range interaction with $\{d:n\} = \{2:1\}$ ($\beta = 0.298$, $\gamma = 1.392$, and $\delta = 5.67$) and 3D Ising model with $\{d:n\} = \{3:1\}$ ($\beta = 0.325$, $\gamma = 1.24$, and $\delta = 4.82$), in correspondence to different classes on the two sides of the magnetic transition. It suggests a change in the nature of magnetic interaction when crossing the transition point, passing from short-range order to long-range order with exchange interaction extending from 2D to 3D on cooling through $T_c$. Furthermore, with the rapid development in the field of 2D materials, we expect our experimental work to stimulate broad interests in reducing bulk Cr$_{0.62}$Te to monolayer sheets and possible spintronic application.

\section*{Acknowledgements}
We thank John Warren for help with SEM measurements. This work was supported by the U.S. DOE-BES, Division of Materials Science and Engineering, under Contract No. DE-SC0012704 (BNL)

\end{document}